\documentclass[preprint]{aastex}
\usepackage{graphics}
\usepackage{graphicx}
\usepackage{multirow}
\usepackage{psfig}
\usepackage{lscape}
\usepackage{rotate}

\begin{document}

\newcommand{\dg}{$^\circ$} 
\newcommand{\spoke}{$\!${\it Q-spoke}}
\newcommand{\XS}{{\sc xspec}}

%%%%%%%%%%%% TITLE %%%%%%%%%%%%%%%%%%%%%%%%%%%%%%%%%%%%%%%%%%%%%%%%
                             % page:title

\title{\bf An X-ray upper limit on the presence of a Neutron Star for the \\ Small Magellanic Cloud and Supernova Remnant 1E0102.2-7219}

\author{M.J. Rutkowski\altaffilmark{1}, E. M. Schlegel\altaffilmark{2}, J. W. Keohane\altaffilmark{3} and R. A. Windhorst\altaffilmark{1}}
\altaffiltext{1}{School of Earth and Space Exploration, Arizona State University, Tempe, AZ 85287, USA}
\altaffiltext{2}{Department of Physics and Astronomy, University of Texas---San Antonio, San Antonio, TX 78249, USA}
\altaffiltext{3}{Department of Physics and Astronomy, Hampden-Sydney College, Hampden-Sydney, VA 23943, USA}
\email{mjrutkow@asu.edu}

%%%%%%%%%%%% ABSTRACT %%%%%%%%%%%%%%%%%%%%%%%%%%%%%%%%%%%%%%%%%%%%%%%%
%page: abstract
\begin{abstract}
%\medskip
We present {\it Chandra X-ray Observatory} archival observations of the
supernova remnant 1E0102.2-7219, a young Oxygen-rich remnant in the
Small Magellanic Cloud.  Combining 28 ObsIDs for 324\,ks of total
exposure time, we present an ACIS image with an unprecedented
signal-to-noise ratio (mean S/N $\simeq$ $\sqrt{S}$ $\sim$ 6; maximum S/N
$>$ 35) .  We search within the remnant, using the source detection
software {\sc wavdetect}, for point sources which may indicate a compact
object.  Despite finding numerous detections of high significance in
both broad and narrow band images of the remnant, we are unable to
satisfactorily distinguish whether these detections correspond to
emission from a compact object.  We also present upper limits to the
luminosity of an obscured compact stellar object which were derived
from an analysis of spectra extracted from the high signal-to-noise
image. We are able to further constrain the characteristics of a
potential neutron star for this remnant with the results of the
analysis presented here, though we cannot confirm the existence of
such an object for this remnant.

\end{abstract}

\section{Introduction}
 The supernova remnant 1E 0102.2-7219 (SNR 1E0102.2-7219, hereafter,
``E0102'') was first identified by \citet{SM81} in a survey of the
Small Magellanic Cloud (SMC) with the {\it Einstein X-Ray Observatory}\@.
While this oxygen-rich remnant resides at a distance of 59\,kpc
\newline\citep{VDB99}, it is far less extincted than many Galactic
remnants due to its high galactic latitude ({\it l} $\sim$
--45$^{\circ}$).  Kinematic studies in the X-ray and optical have
determined E0102 to be a relatively young remnant, with an expansion
age between 1000 and 2000 years \citep[][respectively]{H00,Fink06}\@.
For these reasons, E0102 has been extensively studied across the
electromagnetic spectrum $\!\colon\!$ radio \citep{AB93}, IR
\citep{S05}, optical \citep{TD83, B00, Fink06}, UV \citep{B89}, and
X-ray \citep{H94, G00, R01, F04}\@.

E0102 was likely generated by a Type Ib or Ic
core-collapse supernova event \citep{B00}\@.  As such, the remnant E0102 
would be expected to host some variety of compact object
(i.e., stellar mass black hole or degenerate stellar object).
Furthermore, \citet{AB93} and \citet{Fink06} have considered the
possibility that such a compact object may be influencing the complex
morphology of this remnant.  Though this remnant has been
well studied, few groups have provided observational constraints to
the presence of a compact object for E0102.  In fact, only
\citet{G00} have presented an upper flux limit on a Crab-like pulsar
for this remnant.  In this research, we seek to constrain the X-ray
luminosity of this unidentified compact object.

E0102 is a calibration target for the {\it Chandra X-Ray Observatory}
facility team, and has been observed since {\it Chandra's} 1999
launch.  As a result, the {\it Chandra} Data Archive contains 2.16
million seconds of calibration observations acquired during 215
observing sessions (ObsIDs).  Many of these observations were
obtained to calibrate the off-axis performance of the detector.  We
use archive calibration data exclusively in this research,
specifically those ObsIDs with high spatial resolution acquired
on-axis between 2000 and 2009.  With these combined data, we
present new upper limits to the X-ray flux of a compact
stellar object for the remnant.

%%%%%%%%%%%% OBSERVATION %%%%%%%%%%%%%%%%%%%%%%%%%%%%%%%%%%%%%%%%%%%%%%%%

\section{Observations}\label{sec:archivedcalib}

We used {\it Chandra} archive calibration data of E0102 as observed
with the Advanced CCD Imaging Spectrometer \cite[ACIS; see][for a
  review of ACIS]{G03}\@.  Of the more than 2\,Ms of available {\it
  Chandra} calibration observations of E0102, we select those
observations for which E0102 was imaged on the S3 chip in the ACIS-S
configuration.  We also require that the calibration imaging was
obtained without spectral gratings in the beam.  This selection criteria ensured
that the remnant was not imaged across multiple chips, the
optimal spatial resolution of 1$\farcs$0 was achieved with on-axis
pointings (see \citet{C02} for more details),
and also ensured a uniformity in the data set necessary for the
spectral analysis presented in Section 4.2.  The S3 chip is a back-illuminated CCD
chip, and thus has higher sensitivity to lower energy photons than the
front-illuminated I3 chip, making this chip the favorable one to use
when characterizing any compact stellar object in the remnant with a soft
spectral behavior (e.g. a Cas-A-like neutron star). The ObsIDs that
met these criteria are listed in Table \ref{s3tables.tab}\@.

%%%%%%%%%%%%%%%%%%%%%%%%%%%%%%DATA REDUCTION %%%%%%%%%%%%%%%%%%%%%%%

\section{Data Reduction}\label{sec:datareduction}

Data reduction and processing were completed with the {\it Chandra}
software package, CIAO version 4.1\footnote{see {\tt
    http://cxc.harvard.edu/ciao/}}, and IDL\@.  The ObsIDs were
processed with the CIAO routine {\sc acis\_process\_events} to remove
pixel randomization.  For each ObsID, a lightcurve was generated for a
background region beyond the remnant, to check for possible flaring
events.  Temporal variations above the mean background sky, which
could indicate solar flaring events during the exposure, were found to
be minimal, for each of the ObsIDs selected.  A charge transfer
inefficiency (CTI) correction was applied, to minimize the effects of
radiation contamination which the {\it Chandra} detectors have
experienced in orbit since launch.  Next, the sub-pixel event
re-positioning algorithm {\sc ser}\_{\sc v2}, an IDL procedure, was
applied to each exposure. This routine is designed to improve {\it
  Chandra}'s spatial resolution, by decomposing 2,~3, and 4 pixel
photon splits at the chip interface \cite[see][]{L03}\@.

Using the CIAO script {\sc merge\_all}, the data were re-projected
onto one common frame (ObsID 2843) and stacked to produce a single
exposure with an effective exposure length of approximately
324\,ks. The increase in the signal-to-noise for this combined image
is a factor of $\sim$ 4 -- 6 over the S/N of any individual ObsID.
This stacked image is referred to as the merged image in the text.  This, and all other, images which were used to complete the analysis we present here, at the native pixel scale of the ACIS detector 0$\farcs$49 pixel$^{-1}$.

%%%%%%%%%%%% ANALYSIS %%%%%%%%%%%%%%%%%%%%%%%%%%%%%%%%%%%%%%%%%%%%%%%%

\section{Analysis}

\subsection{Image Analysis}\label{sec:photometric}

\subsubsection{Locating a Compact Stellar Object}\label{sec:testsource}

Simulations and observations alike suggest that core-collapse
supernova events are not likely to be spherically symmetric in their
collapse \citep[see][and references therein]{WW08}\@.  It may be
possible, or even requisite for some varieties of core-collapse
supernovae, that this asymmetric evolution results in a ``pre-natal
kick'' of the compact object revealed by the supernova event
\citep{Fr06, F09}\@.  We include this dynamic factor by defining an
area in which a ``kicked'' stellar object would most likely be
found. This ``test source region'' (hereafter, {\it TSR}) has a
radius, {\it r}, estimated by $r = \frac{v_k}{t_e}$, where $v_k$ is
the ``kick velocity'' and $t_e$ is the expansion age of the remnant.

We use two published expansion ages to constrain $t_e$ for E0102.
From the analysis of the X-ray proper motions of shell ejecta over a
period of approximately 20 years, \cite{H00} determined the age
of E0102 to be equal to $\sim 1$\,kyr. \cite{K01} and
\cite{Fink06} determined from the analysis of medium and broadband Hubble Space Telescope
imaging, that E0102 was significantly older, between $\sim 2.0$ and
2.1\,kyr, respectively.  These estimates to the age of the remnant
agree within the error bars provided by each of the studies.  To
estimate the kick velocity, $v_k$, we adopt the mean two-dimensional (2D) pulsar
velocity of 240 km s$^{-1}$, derived by \cite{HLLK05}
from a proper-motion study of 73 pulsars with characteristic ages less
than 3\,Myr.  With these assumptions, a ``kicked'' stellar remnant
could travel a linear distance between 0.1 and 0.5\,pc.

\begin{figure}[htbp]
%\begin{minipage}{7.11in}
\centerline{
\includegraphics[width=5in,angle=0,scale=1.2]{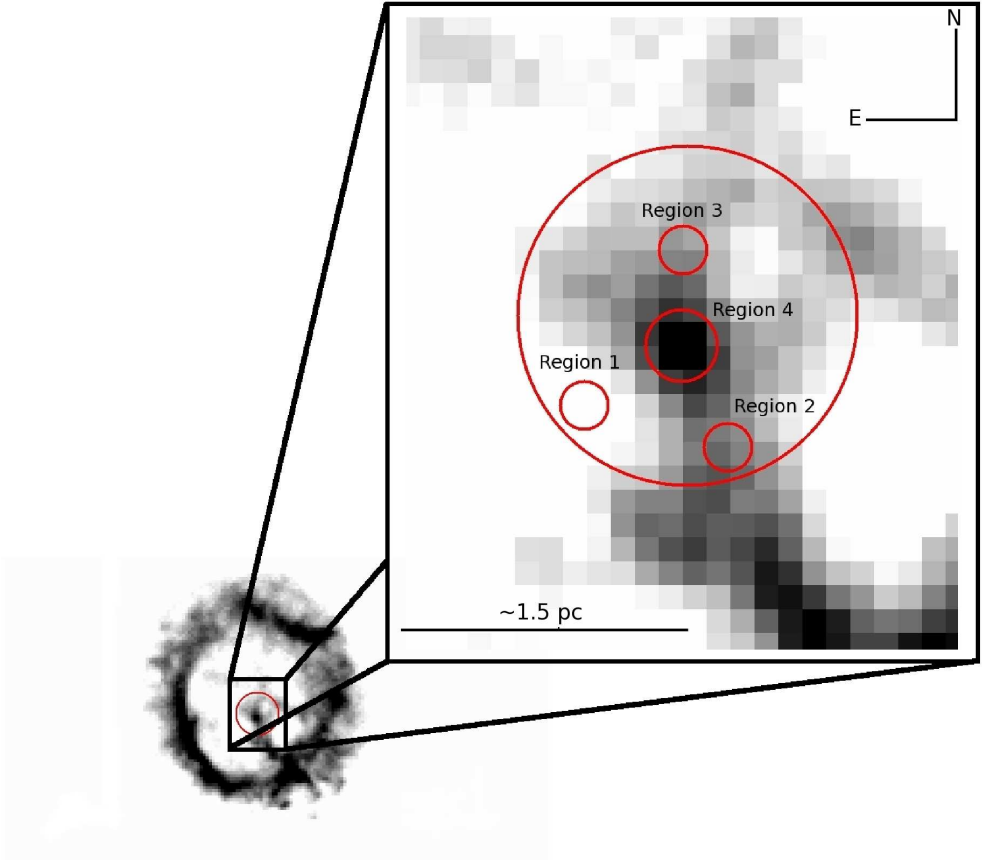}
}
\caption{Remnant 1E0102.2-7219, displayed at the native pixel scale of the ACIS instrument (0\farcs49 pixel$^{-1}$).  Emphasized here is the region in
which a compact stellar object would likely be found (see
Section \ref{sec:testsource}).  The upper limits to the flux from a neutron star in each of these regions are discussed in Section \ref{sec:twosig}\@.}
\label{fig:E0102zoom}
%\end{minipage}
\end{figure}

We assume the proper motion center (R.A. $\!\colon\!$
  01${\rm^h}$04${^m}$02${^s}.05$, decl. $\!\colon\!$
  -72${^{\circ}}$01'54\farcs99 (J2000)) determined in \citet{Fink06}
for the center of our {\it TSR}\@.  The physical distance between this
center and the main ejecta shell of the remnant is approximately
6\,pc. Thus, a ``kicked'' stellar remnant with the aforementioned
properties could not have reached the remnant shell in the
expansion lifetime.  With the assumption of the \cite{Fink06} center,
we define the {\it TSR} as a generous circular area with a radius
$\simeq$ 1.2\,pc, which we overlay on the map of the remnant in Figure
\ref{fig:E0102zoom}\@.

\subsubsection{Galactic Analogs to an E0102 Central Compact Stellar Object}\label{sec:twosig}

In the photometric analysis we perform here, we restrict our search
for a stellar object in the remnant to two well-studied potential
analogs $\!\colon\!$ the central compact object (CCO) in
Cassiopeia-A \citep[see][and references therein]{C01,F06,P09} and
the Crab pulsar \citep[see][and references therein]{H08}\@.

\cite{G00} provide an upper limit to the X-ray luminosity of
9$\times$10$^{33}\,$erg~s$^{-1}$ for a Crab-like pulsar in the
remnant.  This limit was determined from a 9.7\,ks observation of the
remnant with the {\it Chandra} ACIS instrument.  Further, the
spectroscopic study of this remnant by \cite{F04} implies that if a
Crab-like compact stellar object is present, then it must be faint at X-ray
wavelengths.

If the E0102 supernova event generated a Crab-like pulsar, then this
object's spectral behavior would be well described by a power-law,
with an approximately constant photon index ($\Gamma$) in the {\it
  Chandra} bandpass.  For the Crab pulsar, $\Gamma\simeq$ 1.6 - 1.8,
but the spectrum of the Crab remnant is softer and better fit with
$\Gamma\simeq$ 2.0 \citep{W04,S06}\@.  At low energies (energy $\sim$
0.2 - 2.0 keV), the relative strength of a thermal emission component arising
from the remnant plasma is likely to overwhelm the contribution from
the Crab-like pulsar component. But, at higher energies, the relative
strength of the Crab-like component would increase, possibly to the
extent that the compact object could dominate the emission in the
center of the remnant, as was observed for the pulsar wind nebula
associated with SNR G292+1.8 \citep{H01}\@.

To identify such a trend for this remnant, we first processed the
merged image to isolate high and low energy emission between 0.5 and
10.0\,keV within energy bins whose widths were selected to
include the following $\!\colon\!$ known complexes of metal emission-lines of oxygen
(\ion{O}{7} triplet $\!\colon\!$ 0.46 -- 0.6\,keV and \ion{O}{8}
Ly$\alpha$ $\colon$ 0.61 -- 0.72\,keV), neon (\ion{Ne}{9} triplet
$\colon$ 0.86 -- 0.98\,keV and \ion{Ne}{10} Ly$\alpha$ $\!\colon\!$ 0.98 --
1.10\,keV), magnesium (\ion{Mg}{11} triplet $\!\colon\!$ 1.32 -- 1.4\,keV
and \ion{Mg}{12} Ly$\alpha$ $\!\colon\!$ 1.45 -- 1.53\,keV); those energies
which provide upper and lower energy bounds for each of the previous
line complexes (which could be used to identify continuum sources),
i.e., a lower bound to oxygen (``Bin 1'' $\!\colon\!$ energy $=$
0.20--0.40\,keV), a lower bound to neon and upper bound to oxygen
(``Bin 2'' $\!\colon\!$ energy $=$ 0.75--0.85) and a lower bound to
magnesium and upper bound to neon (``Bin 3'' $\!\colon\!$
energy $=$ 1.15--1.275\,keV); and the hard (2.0--8.0\,keV) continuum.
We visually inspected the resulting X-ray spatial flux maps for each
of these energy bins, but did not identify a compact object. A
Crab-like pulsar should, in addition, be detected across a long
wavelength baseline, but the aforementioned studies of E0102 in the
radio, optical, and infrared wavelengths have not detected a neutron
star. For these reasons, it is unlikely that a Crab-like
neutron star is present in this remnant, and for the photometric
analysis in this section we consider exclusively the presence of a
Cas-A-like CCO.

\cite{C01} combined $\sim 6$\,ks of {\it Chandra} ACIS (S3 chip)
observations of Cas-A, determined the location of the compact stellar object,
and modeled the luminosity of a neutron star for the remnant.  The
energy of the maximum count rate for this object peaked at 1.6~$\pm
 0.2$ keV and was well-fit with an ideal blackbody energy distribution
(kT$_{\mbox{\tiny eff}}$ = 0.5\,keV), assuming a column density
$N_{\rm H}$ = 1.1 $\times$ 10$^{20}$ $\mbox{cm}^{-2} $.  The X-ray luminosity
of the Cas-A CCO is equal to $L_{0.1-10\mbox{\footnotesize keV}}$ =
2.0$^{+0.8}_{-0.6}$ $\times$ 10$^{33}$ $\mbox{erg}$ s$^{-1}$. Artificially moving
this object to the distance of E0102 (59\,kpc), the unabsorbed flux in
the same bandpass equals 4.7$\times$10$^{-15}$
$\mbox{erg}$~s$^{-1}$~$\mbox{cm}^2$, or equivalently a count rate
equal to 7.15$\times$10$^{-4}$ counts s$^{-1}$.  In 324\,ks, this count rate is
equivalent to 221 counts.  We cannot visually identify a compact object,
with this count rate, within the {\it TSR}\@.

An examination of the emission from the remnant within the {\it TSR}
does reveal a strong gradient in the surface brightness, though.  In
Figure \ref{fig:E0102zoom}, Region 1 contains an area within which the
emission is of the lowest mean surface brightness within the {\it
  TSR}\@.  In this 324\,ks exposure, the mean count number within this
class of region equals 117.  The \spoke, the hot extended plasma
feature extending from the proper motion center to the shell of the
remnant, falls within the {\it TSR} and defines a ``medium'' class of
surface brightness regions with average count rates $\sim$ 5 times
greater than the low surface brightness regions previously discussed.
Regions 2 (mean count $\approx$ 521) and 3 (mean count $\approx$ 459)
are regions representative of this class of emission.  Within a 2
$\times$ 2 square pixel region (center $\!\colon\!$ R.A. -
  01${\rm^h}$04${^m}$02${^s}.015$, decl.-
  -72${^{\circ}}$01\farcm55\farcs82 (J2000)), the unique location of
peak emission within the {\it TSR} defines a third class of surface
brightness. The mean count value for this region, Region 4, equals
$\sim$ 1157 counts.  This variation in surface brightness could imply
that a point source exists within the remnant but that the remnant's
hot plasma ``masks'' the emission that arises from the compact object.

If we assume that a neutron star with a spectral behavior similar to
the Cas-A CCO is in fact present in this remnant, but is concealed or
masked by the X-ray bright plasma along the line of sight, then the
observed count rates can be used to produce upper limits to the
luminosity of such a neutron star.

We assume a line of sight SMC H\small{I} Column Density\footnote{The Milky Way
  Column Density, estimated with the CXC software package COLDEN
  {an online tool accessible at {\tt http://cxc.harvard.edu/toolkit/colden.jsp}}, is
  low relative to the SMC absorption, with $N_{\rm H}$ = 6.57 $\times$
  10$^{20}$ cm$^{-2}$}, $N_{\rm H}$ = 5.82 $\times$ 10$^{21}$
cm$^{-2}$ from \cite{S99}\@.  With PIMMS\footnote{Available at {\tt
    http://cxc.harvard.edu/toolkit/pimms.jsp}}, we then calculate the
upper limit to the flux, and derive the luminosity, for a Cas-A
neutron star with either a blackbody or power-law spectral energy
distribution.  For these calculations, we set the blackbody
kT$_{\mbox{\tiny eff}}$ parameter and the power-law photon index
$\Gamma$ equal to the similar best-fit parameters derived by
\cite{C01}\@.  We present these upper limits in Table
\ref{tab:powerlawest}\@. In this table, the 5$\sigma$ rms sky noise
(in counts) is never greater than $\sim 3$, which implies less than
$\sim 3\%$ of the counts for each of the regions in Table
\ref{tab:powerlawest} are detector, or sky, background not arising from
the remnant.  We do not attempt here to estimate the
background ``noise'', within each region, which arises from the remnant
plasma directly (see Section \ref{sec:specanalysis}).

A compact object is only one of many components contributing to total
flux observed within a region of interest, and within the {\it TSR}
(particularly within Region 4), the contribution from this source may
be small relative to other components (background and sky ``noise'',
detector noise, detector response, absorbers along the line of sight,
etc.).  Nevertheless, for regions which are not coincident with the
\spoke, the ``zeroth-order'' estimates to the upper limits to the
flux from a compact object in these regions may be reasonable, as the
contribution of a compact object in one of these regions cannot, by
definition, exceed the observed flux.  But, the {\it TSR} includes
bright emission from the \spoke, and the surface brightness variations
indicate that the contribution of the remnant plasma is variable with
position.  Thus, quantifying the contribution of a compact object to the
total observed flux for each region requires a more rigorous estimation
of the contribution of the background to the total observed flux.  In
the following sections, we present the results of two methods which
provide a more robust determination of the relative strength of a
compact object within the {\it TSR}\@.

\subsubsection{Point-source Identification}\label{sec:pointsources}

To date no study of E0102 has reported the detection of a compact
object in physical association with the remnant. But, previous
investigations of E0102 were limited in their sensitivity and spatial
resolution.  Combining the extensive archived observations of this
remnant, we have produced a high signal-to-noise image of the remnant,
at the superior X-ray spatial resolution of the {\it Chandra} ACIS
instrument suite, which represents the best opportunity to identify a
compact object within the {\it TSR}, particularly in the vicinity of
Region 4.

In the previous section, it was found that complex surface brightness
variations (with radius, measured from the center of the remnant)
prevent any conclusive visual identification of a compact object.  In
this section, we search for a point-source object in the image using
the CIAO {\sc wavdetect} script, which is the optimal source detection
software package for {\it Chandra} X-ray images.  This software
iteratively identifies source objects by correlating image pixels with
``Mexican Hat'' wavelets, for a selection of user-defined wavelet
scales\citep[for a description of the software, see][]{F02}\@.  We
apply {\sc wavdetect} for source detection in seven images of the
remnant$\colon$ {\it Broad} (energy $=$ 0.5--7.0 keV), {\it Medium}
(energy $=$ 1.2--2.0 keV), {\it Soft} (energy $=$ 0.5--1.2\,keV), {\it
  Hard} (energy $=$ 2.0--7.0 keV)\footnote{These first four energy
  bands are directly comparable to Chandra Source Catalogue bands.},
and the three energy bands discussed in Section \ref{sec:twosig} that
selected continuum emission between strong line complexes.  For each
of these images, {\sc wavdetect} was used to identify sources on
``Fine'' (scales = ``1 2'') and ``Coarse'' (scales = ``4 8 16'')
scale, with {\it sigthresh}\footnote{{\it sigthresh} defines the
  threshold at which ``cleaned'' pixels are considered to be members
  of a source detection} set equal to 4$\times$10$^{-5}$.  We ran each
{\sc wavdetect} search twice, with and without an effective exposure
map, and found the same number of detections within the {\it TSR}\@.

\begin{figure}
%\begin{minipage}{7.11in}
\centerline{
\includegraphics[width=5in,angle=0,scale=0.6]{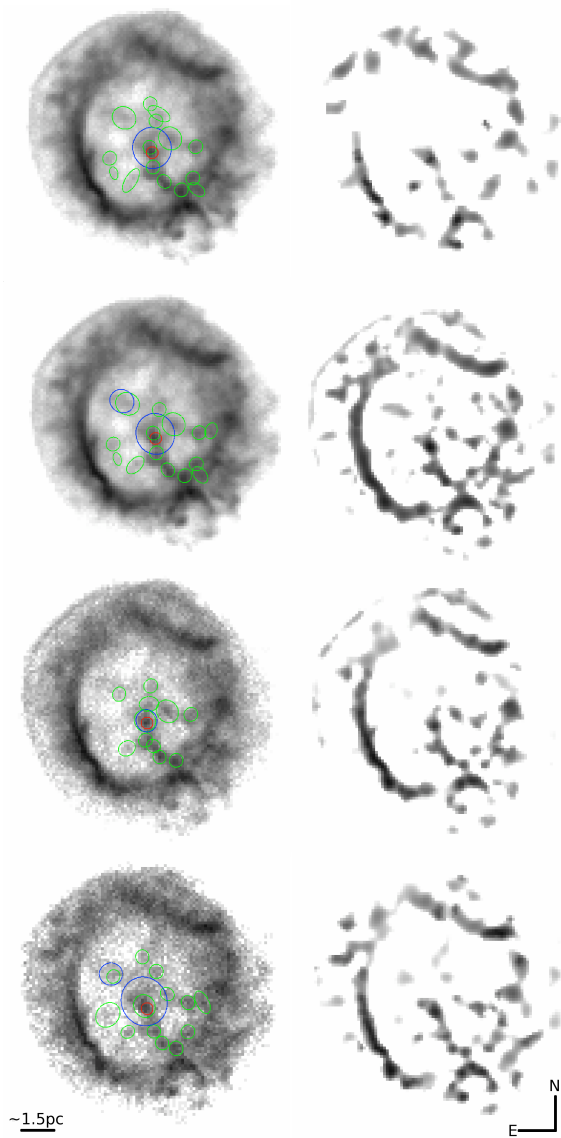}
}
\caption{ Displayed (from top to bottom left hand column) are the {\it broad} and {\it hard} broadband images and ``Bin 2'' and ``Bin 3'' narrowband images of E0102 at the ACIS native pixel scale (0$\farcs$49 pixel$^{-1}$) in gray scale.  The online edition of this figure is provided in color. Overplotted on each image are the sources detected within 10$\farcs$ of Region 4 (see Figure 3) with CIAO package {\sc wavdetect} for the ``Fine''-scale (in green) and ``Coarse''-scale (in blue) wavdetect searches. A red region is also plotted to indicate the center of Region 4. Displayed in the right column is the image file produced by {\sc wavdetect} which represents the remnant with the estimate of background and sky noise subtracted. Definitions of the energy bands on which these images were binned the {\sc wavdetect} searches performed and the results and discussion derived from these source detections can be found in Sections 4.1.2 and 4.1.3.}
\label{fig:sources}
%\end{minipage}
\end{figure}

{\sc wavdetect} searches in each of these images identified a source within Region 4,
and the detection ellipse for each of these sources included part (or
all) of this region.  In Figure \ref{fig:sources}, we overplot the
source detections which were identified within 10$\farcs$0 of the
expansion center for the {\it Broad}, {\it Hard}, ``Bin 2'' and
``Bin 3'' continuum images.  In addition, we include in Figure
\ref{fig:sources} the ``image file'', an output from the {\sc
  wavdetect} routine that is a representation of the data with
background and noise subtracted.  We are reluctant to conclude that these source
detections are associated with a compact object, though, and believe that
these results reinforce, quantitatively, the preliminary conclusion
presented in Section \ref{sec:twosig} that a compact object may be
present in or near to the brightest pixel region (Region 4) but masked, for the following reasons.

First, the {\sc wavdetect} results confirm that the \spoke~is a real
feature, contributes significantly to surface brightness variation
across the center of the remnant, runs continuously between the center
and outer edge (from the observer's perspective) of the remnant as
illustrated in the ``image file'', and the brightest pixel region
remains ``masked'' in the \spoke~in the ``image file''.

{\sc wavdetect} provides a variety of photometric measurements which
can be used to constrain the extended versus\,compact nature of detected
objects (e.g., PSFRATIO\footnote{{\sc wavdetect} calculates this by
  dividing the estimated ``radius" of the source cell to {\it
    PSF\_SIZE}, an estimate of the point-spread function (PSF) calculated at the location of
  the detection.  For further information, please see the CIAO {\it Detect}
  Reference Guide at {\tt
    http://cxc.harvard.edu/ciao/download/doc/detect\_manual/wav\_ref.html}})
and an estimate to the statistical significance of each detection
(SRC\_SIGNIFICANCE\footnote{{\sc wavdetect} provides an estimate to
  the significance of the detection with respect to the calculated
  background for the source detection region, but should not be taken
  at face value.  This statistic is defined in the CIAO Detect
  Reference Guide and is calculated by dividing the net counts by the
  ``Gehrels error'', $\sigma_G = 1 + \sqrt{BKG_{\rm COUNTS} +
    0.75}$.}). When the source detections nearest to the brightest
pixel region are considered individually, these statistical
measurements could suggest possible point source detections,
particularly in the narrow band images.  Considered in the context of
the full set of detections, though, the sources detected in the
proximity of Region 4 are indistinguishable from other detections
within the remnant, most of which are associated with hot gas
background emission that could not be reduced from the image file by
the {\sc wavdetect} software.  Furthermore, the lack of nearby point
sources or double-peaked source detections make it difficult to
interpret these results. In Table \ref{tab:wavdetect} we provide
pertinent statistics for the set of detections from each of the {\sc
  wavdetect} searches; here, we limit the set to those which have
detection ellipse centers ($\alpha,\delta$) interior to the outer
shell of the remnant.

  Lastly, each image's source detection ellipse most closely matching
  to, or including Region 4, in the ``Fine''-scale {\sc wavdetect}
  search is found to be offset from the center of Region 4 by less
  than 1$\farcs$0.  But the size of the detection ellipse
  significantly increases at the ``Coarse'' scale.  These latter two
  results likely indicate the difficulty which {\sc wavdetect} has in
  adequately discriminating between sources and unresolved background
  components, or knots of hot plasma and ``masked'' compact objects.

Though the {\sc wavdetect} search results provide more point source
candidates within the {\it TSR}, we believe that any photometric
method of estimating the relative strength of a compact object
``masked'' by the \spoke~is inadequate and now turn to a spectral
analysis for characterizing the background emission.

\subsection{Spectral Analysis}\label{sec:specanalysis}

In the previous sections, we were limited in deriving firm upper
limits on the detection and characterization of a compact object by
the presence of a high surface brightness emission component that
likely arises from the remnant plasma.  In this section, we perform a
spectral analysis designed to quantify the contribution of multiple
background components present within an expanded set of regions within
or near to the {\it TSR}\@.  Only by fitting and reducing out the
contribution of these components in the regions of interest will we be
able to conclude firm upper limits for compact objects which may be
masked by the \spoke.

It is necessary to note that E0102 is used by the {\it Chandra}
facility as a spectral calibration source, because it is a relatively
constant and bright X-ray source, especially at energies between 0.55
and 0.7\,keV corresponding to the emission-lines of \ion{O}{7} and
\ion{O}{8}\@.  As such, it is an optimal target for use in calibrating
the response of the ACIS CCDs and the variability of these CCDs
response in time (see, e.g., \cite{P08}).  Imaging of the remnant for
calibration purposes does not preclude us from searching for multiple
additional spectral components, though. 
                                   
\begin{figure}
%\begin{minipage}{7.11in}
\centerline{
\includegraphics[width=5in,angle=0,scale=0.85]{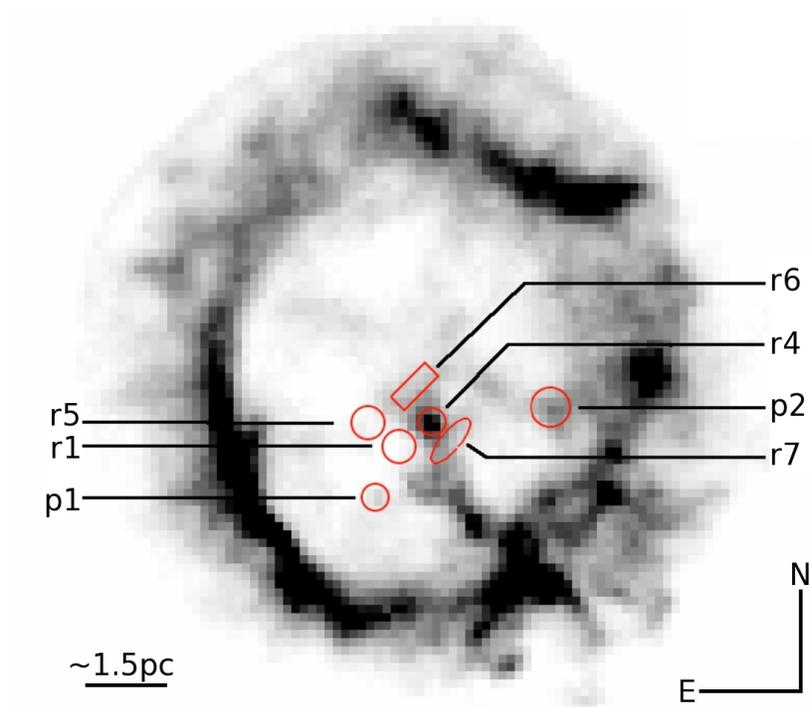}
}
\caption{Overlaid on this image of the remnant are those regions for which a spectral analysis was performed; we refer the reader to Section \ref{sec:specanalysis} for complete details and results of this analysis.  The remnant is displayed at the native pixel scale of the ACIS instrument (0\farcs49 pix$^{-1}$).}
\label{fig:E0102zoom_spec}
%\end{minipage}
\end{figure}

Using the CIAO script {\sc specextract}, we extracted spectra
for each of the regions discussed in Section \ref{sec:photometric}
(see Figure \ref{fig:E0102zoom}) and an additional six regions of
varied size, shape, and location.  We select these regions to better
characterize the background emission within the remnant, not just
within the \spoke~and {\it TSR}\@. In Figure \ref{fig:E0102zoom_spec}, we
provide a map of the regions, interior to the remnant shell, for which
spectra were extracted.  Not displayed in Figure
\ref{fig:E0102zoom_spec} is the background region (area $\simeq$
27,000 pixel$^2$) which was used to characterize the X-ray sky
background not originating from the remnant itself. These regions were
selected from areas of the chip where the individual ObsID exposures
were, when stacked, spatially coincident at distance well beyond the
remnant.  For legibility, Figure \ref{fig:E0102zoom_spec} does not
include Regions 2 and 3.

In this section, we are concerned exclusively with the detection of a
point source object by its spectral signature. In theory, the
detection of any such object is possible only by identifying and
reducing, from the spectral profile extracted for a particular region
of interest, all emission which does not arise from the point source
itself so that the flux from the compact object can be properly measured.  We
assumed that this background arose from one of three sources$\colon$
the X-ray background sky arising from astrophysical sources
independent of the remnant as well as instrumental background
particular to the ACIS detector (i.e., ``chip noise''), which we will
refer to here as ``Component A'' background, and diffuse emission
arising from E0102 ejecta, the intensity of which varies with the
position in the remnant, which we will refer to as ``Component B'' background.

To calibrate the emission arising from the X-ray sky (e.g., diffuse
galactic Milky Way and SMC emission), and the
background arising from the ACIS detector electronics (e.g., the
well-understood fluorescence lines of Si K$\alpha$, Al K$\alpha$,
etc.; Baganoff 1999), we first analyzed a
spectrum extracted from a uniform region (the aforementioned
``background region'') located approximately 1' from the outermost
radial extent of the remnant.  Using the {\sc xspec}\footnote{see {\tt
    http://heasarc.gsfc.nasa.gov/docs/xanadu/xspec/}} software
package, we then modeled the broadband spectral behavior of this
``Component A'' emission using multiple power-law and Gaussian
components to represent this diffuse X-ray sky emission and
lines. With 583 degrees of freedom in the model fit to ``Component A''
background emission, the reduced minimized $\chi^2$ statistic for the
best-fit model is equal to 1.03. For this analysis, and throughout
this section, we consider the spectral profile in two energy bins,
``soft'' X-rays (0.5 -- 2.0\,keV) and ``hard'' X-rays (2.0 -- 8.0\,keV
).

``Component A'' background emission is assumed to be uniformly present
and of similar intensity within each of the regions for which we
extracted spectra.  This assumption does not hold for ``Component B''
emission which arises from the remnant itself.  To fit this class of
background emission, we must consider each of the spectra
individually.  Because our goal is to identify a spectrally distinct
compact object in this remnant, we aim to characterize the diffuse
plasma spectrum rather than to physically understand it.  Therefore,
we fit an ad hoc model to the spectrum for each of the regions.
Again, we used the \XS~package to model this background
emission, fitting the ``Component B'' X-ray emission (which was
dominated, not unexpectedly, in the soft band by the strong oxygen and
neon emission-line complexes) with a system of Gaussian line profiles,
and also with a variable APEC model \citep[see][]{S01}\@.  We
acknowledge here that the spectral resolution is insufficient for
distinguishing between spectral lines and continuum emission between
those lines; thus the fits to these sources of background emission are
model dependent.  Therefore, in the following analysis, the upper
limits to the continuum emission from any putative neutron star are
implicitly model dependent.

\begin{figure}
%\begin{minipage}{7.11in}
\centerline{
\includegraphics[width=5in,scale=0.35,angle=270]{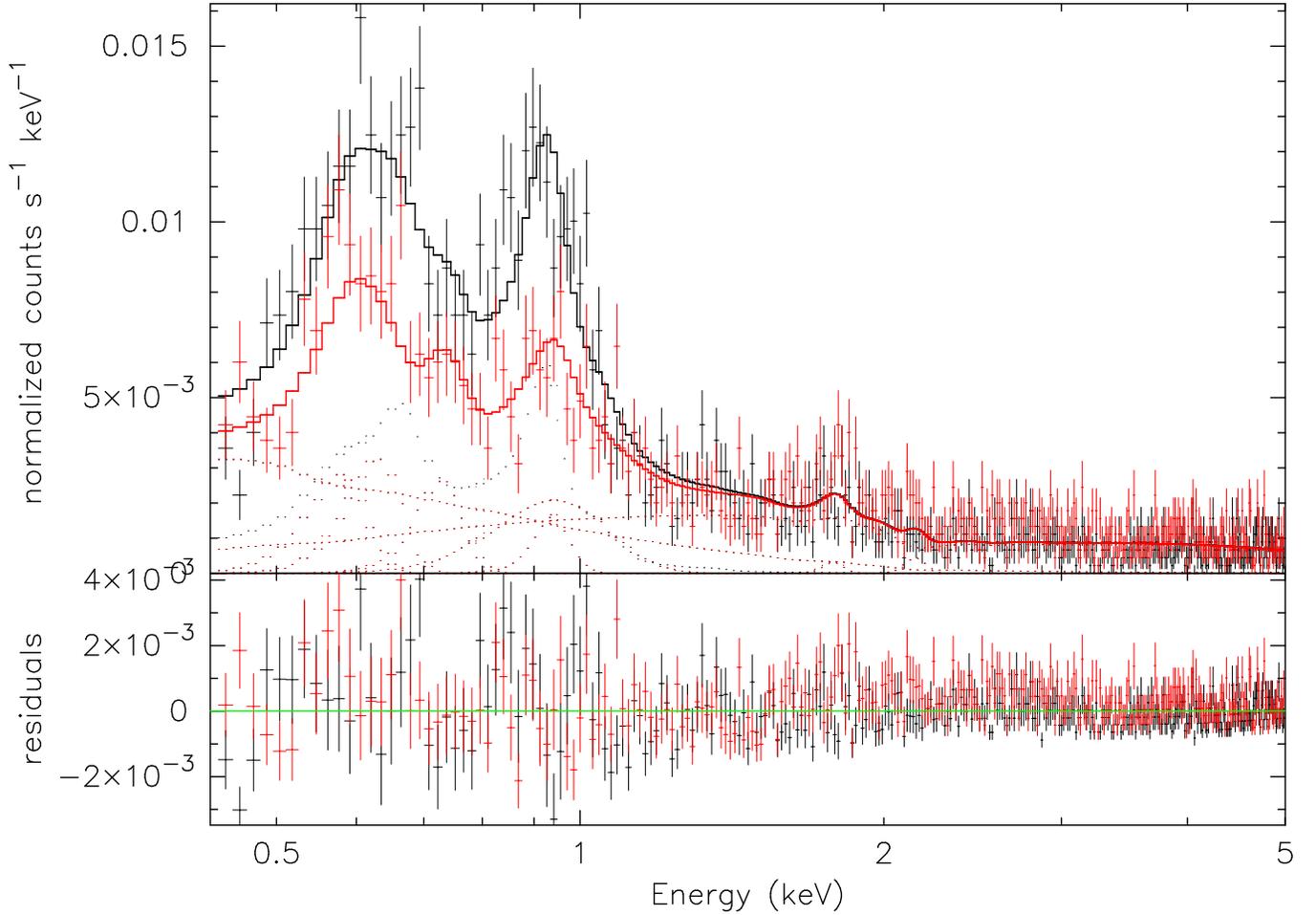}
}
\caption{``Soft'' imaging spectrum extracted from the brightest
  pixel region in the remnant (Region 4, see Figure
  \ref{fig:E0102zoom_spec}) is plotted with observational error bars
  in black.  Overplotted (in red and black) are the best-fit models to
  the background emission arising from all sources.  Here, the nebular
  line emission complexes (``Component B'' emission) have been modeled
  with a variable APEC thermal plasma model
  in {\sc xspec}.  Please see Section \ref{sec:specanalysis} for
  details on the sources of background emission and the results of the
  spectral analysis performed for each of regions identified in
  Figures \ref{fig:E0102zoom} and \ref{fig:E0102zoom_spec}.}
\label{fig:background}
%\end{minipage}
\end{figure}

\begin{figure}
%\begin{minipage}{7.11in}
\centerline{
\includegraphics[width=5in,scale=0.35,angle=270]{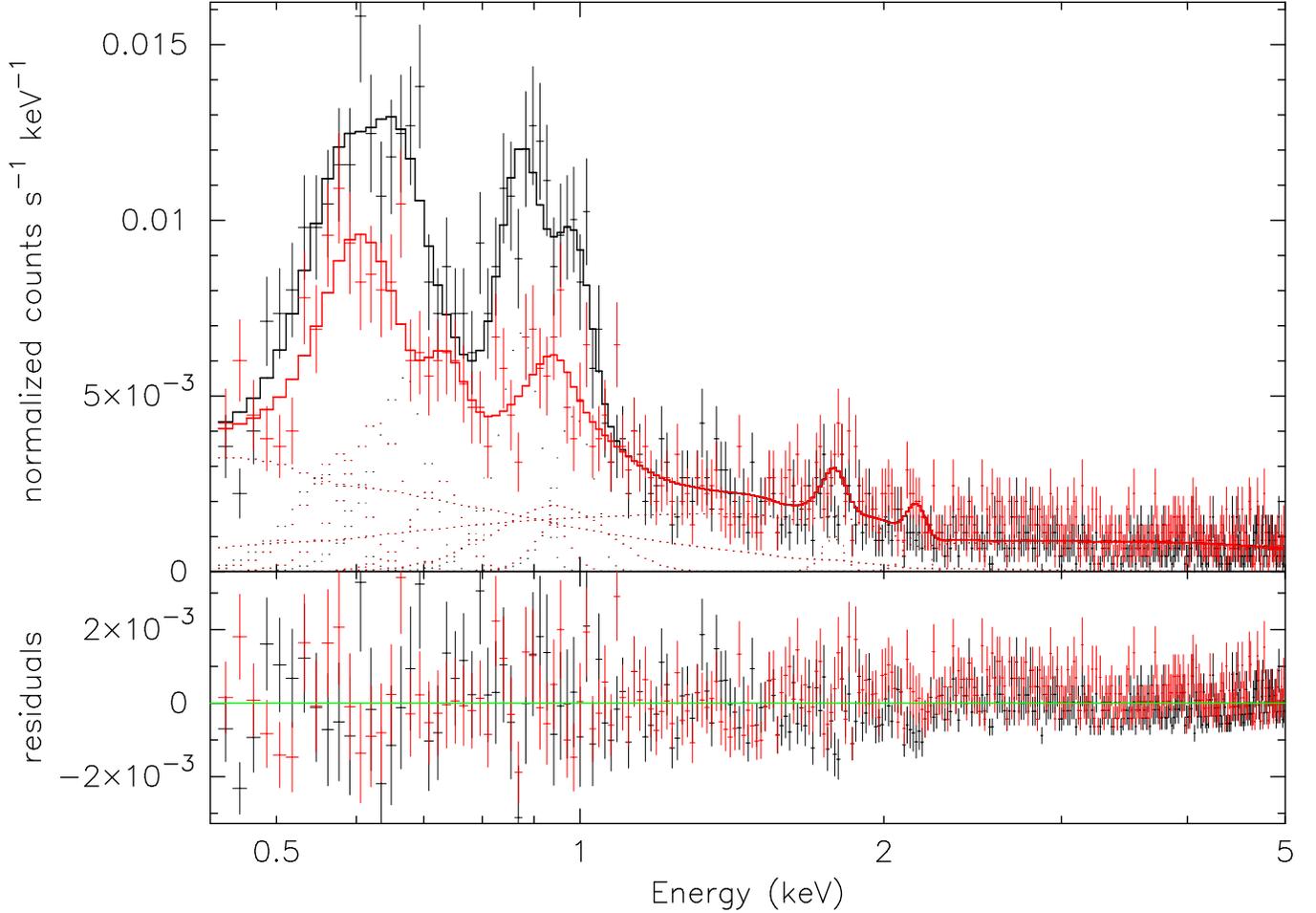}
}
\caption{Same as Figure \ref{fig:background}, the ``soft'' imaging
  spectrum extracted from the brightest pixel region in the remnant
  (Region 4, see Figure \ref{fig:E0102zoom_spec}) is provided here, with
 model fits to the background emission overplotted. Here, the hot gas
  background component arising from the remnant has been modeled with
  Gaussian spectral lines in {\sc xspec}.  Please see Section \ref{sec:specanalysis} for more
  details.}
\label{fig:background2}
%\end{minipage}
\end{figure}

For each region, we combine both background models generated to
produce a unified background model which we then fit to each of the
observed spectra.  In Figure \ref{fig:background}, we present the
unified background model in its component form. In this figure, black
data points with error bars represent the observed imaging spectrum
extracted from, in this example, the brightest pixel region (Region 4).
The ``Component A'' background model fit is overplotted in red$\colon$
the red solid curve represents the sum of the individual model
components (red dotted line) fit to the red observed data points
(plotted with error bars) of the extracted spectrum from the
background region.  The unified background model, which includes the
region-specific ``Component B'' background emission, is overplotted in
black; in this figure the Component B emission model was generated
using a variable APEC model and power-law components.  For comparison,
we provide in Figure \ref{fig:background2} the background model fit,
but with ``Component B'' emission-line complexes modeled with a system
of Gaussian spectral lines.  Each model reproduces the observations
well, within the given errors, but in general the vAPEC models have
smaller residuals.

The best-fit background model was determined for each of the extracted
spectra by iteratively minimizing the $\chi^{2}$ statistic. The
``Minimum $\chi^2$'' value for each best-fit unified background model
is provided in Table \ref{tab:fluxtab}\@.  Each model, for both the
vAPEC and Gaussian fits, was defined with 577 degrees of freedom, thus
the reduced $\chi^2$ for each of the background fits was $\sim$ 1.0
(e.g. Region 4, $\chi^2_{\nu} = 1.07$).  As ``Component A'' background
emission was fixed in its contribution to the spectral energy
distribution extracted for each region, this figure of merit
essentially measures the goodness of the fit of ``Component B''
background emission, which arises predominantly from nebular line
emission.

Having sufficiently accounted for all sources of background emission,
we can then derive an upper limit to the presence of a compact object
within each of the regions. To derive these upper limits, we assume
that a compact object is a neutron star and is well described by
either a power-law or a blackbody model (hereafter, referred to as the
``point-source model'').

 We define the point-source model as the product of a normalization
 constant and the specific functional form; for example, the power-law
 model point-source model function is $F_{E} = \beta \times
 \mbox{E}^{-\Gamma}$.  Combining the best-fit unified background
 model, now fixed for each region by the previous analysis, with the
 point-source model, we define a two-component spectral model which we
 then re-fit to the observed spectrum for each region by only varying
 the magnitude of the normalization constant \footnote{We fix $\Gamma$
   (or kT, in the case of the Blackbody point source model) for each
   fit.}\@.  We increase the strength of the point-source component in
 this two-component model only until $ \Delta \chi^2 = \mbox{min.}
 \chi^2 + 6.63 $ is no longer satisfied.  Note that 6.63 is the
 chi-square difference constant implying an observation is significant
 at the 99\% (3 $\sigma$) confidence level for a system modeled with
 1 degree of freedom. Thus, with this analysis we have measured the
 upper limit to the observed flux from a compact stellar object, above which
 such an object would have been detected in this data.

In Table \ref{tab:fluxtab}\@ we present the results of this analysis,
which we completed for three power-law models of a compact stellar object
defined with $\Gamma$ set equal to 1, 3, and 5,
respectively.  This range of indices spans the observable parameter
space for the Cas-A CCO \citep[power-law fit, see][]{P09} and pulsars
(i.e., the Crab pulsar).  In Columns 2 -- 6 of Table \ref{tab:fluxtab}
we present the flux, equivalent luminosity, and the minimum $\chi^2$
statistic we determine from the best-fit unified background model to
the spectrum extracted for each regions.  In Columns 8--13 we present
the upper limits to a neutron star for E0102. Specifically,
in Columns 8 and 11, we present the flux difference between the two component best-fit model and the unified background fit for the region, for the soft and
hard energy bands respectively.  Positive values in both the soft and
hard energy bands indicate that the normalization parameter that was found to
satisfy the statistical criterion (defined previously) in one band,
simultaneously satisfied the criterion in the other band.  Or,
similarly, the point-source model flux in both bands was sufficiently
greater than the unified background model to be detected, minimally at
99\% confidence.  Negative or zero values in Columns 8 and 11 should
be interpreted as non-detections at the 99\% confidence level.  These
non-positive values can be attributed to a lack of discriminatory
power in the models for differentiating between remnant hot gas
emission and the point source component, within the observational
errors, at the confidence level defined by the statistical
criterion. The remnant is dominated by soft emission; thus these
non-detections typically arise for the softest ($\Gamma$ = 5) point
source component model fits. We could produce upper limits for these
models if we instead chose to enforce a stricter confidence level,
e.g. 99.5\%, which would require us to redefine the statistical
criterion (in this example, $ \Delta \chi^2 = \mbox{min.}  \chi^2 +
7.88 $).  Lastly, in Columns 9 and 12 we present the flux
upper limits, and in Columns 10 and 13 we present the equivalent
luminosity upper limits for the stellar object.

Because the brightest pixel region (Region 4 in Figure 2) has a
distinct photometric signature (see Sections \ref{sec:twosig} and \ref{sec:pointsources}), we perform a similar
analysis to that which was outlined above for this region, but also
explore the possibility that the point source component within the two
component model is best described by a blackbody model.  We present
this analysis for both the power-law ($\Gamma$ = 1, 3, and 5) and
Blackbody (kT$_{\mbox{\tiny eff}}$ = 0.25, 0.5, and 1.0) model point
source components in Table \ref{tab:region4flux}\@.

%%%%%%%%%%%%%%%%%%%%%%%%%%%%%%%%%%%%%%%%%%%%%%%%%%%%%%%%%%%%%%%%%%%%%%%%%%%%%%%%

\section{Discussion}

We interpret the non-detection of a compact stellar object in the
remnant E0102 within the context of three physical scenarios.  In the
first of these scenarios, the non-detection of a compact stellar
object for E0102 implies that such an object simply does not exist.
There are two plausible scenarios to consider when addressing this
possibility.

First, a stellar compact object may have been destroyed during the
supernova event.  If E0102 represents the end stage of a Type Ia
supernova event, then the progenitor white dwarf(s) would have likely
been destroyed \citep{WW86}\@.  But, it is difficult to reconcile this
formation scenario with the evidence presented by
\citet{B00} and \citet{C05} which would support a Type Ib/c or IIb/L
class supernova event for the E0102 progenitor.

The theory of Type Ib/c and Type II supernova predicts that compact
objects are generated during the supernova event, and compact stellar
objects have been observed in physical association with $\sim$ 25\% of
Galactic remnants \citep[see {\it Green's Catalogue};][]{G06}\@.  But, in
addition to these neutron-degenerate stellar objects, core-collapse
supernovae of sufficiently massive ($>$ 20$M_{\odot}$) progenitors
may generate black holes \citep[see][for a review]{S09}\@.  At
least two arguments exist in support of the black hole scenario, and
though they are circumstantial they should be addressed in future work
on the remnant.  First, \citet{Fink06} suggest that E0102's
progenitor star may have been a massive ($>$50 M$_{\odot}$) Wolf-Rayet
star, based on the consideration of the local environment of the
remnant.  Second, to explain a similar remnant morphology in the SNR
W49B, \citet{K06} suggest a possible jet-driven explosion mechanism \citep[a
``collapsar'' model;][]{M00}, which would generate a black hole.

A second interpretation of the non-detection of a compact stellar
object in the remnant is that a neutron star in this remnant is in
fact present, but is both faint in its emission and obscured by the
\spoke.  This possibility is motivated by the imaging and analysis
presented in Sections \ref{sec:twosig}, \ref{sec:pointsources}, and \ref{sec:specanalysis}\@.

A final interpretation of the non-detection of a compact stellar
object is systemic, such that the estimate to N$_{\rm H}$ is too low.
Thus, the flux from a compact object in this remnant is absorbed by
the interstellar medium (ISM) along the line of sight. At a distance of 59\,kpc, the Cas-A
CCO would have a detectable signal; in this 324\,ks ACIS image, this
signal would be equivalent to $\simeq$ 240 counts between the energies
of 0.1 and 10\,keV.  Such a source would stand well above the noise in
this observation.  But, the average number count within that portion
of the {\it TSR} unobscured by the \spoke~is equal to $\sim$ 140 in
the 324\,ks exposure.  By measuring the relative strength of the
background in an image filtered for energies greater than $\sim$ 10
keV (above which the effective area of {\it Chandra} $+$ ACIS is
nominally zero with respect to the known background for the S3 chip),
an estimate to the cosmic ray background count rate can be made.  With
this rate appropriately scaled to the exposure length and area of this
region, it can be shown that only a small fraction of these counts
($<<$ 1 \%) can be attributed to cosmic rays. Thus, to reduce the
Cas-A-like neutron star flux to this observed level would require a
column density along the line of sight $\sim$12 times higher than we
assumed. This is not reasonable given the errors on N$_{\rm H}$, and
we conclude that gas along the line of sight is not likely to obscure
the flux from a compact object in this remnant.

The non-detection of a compact object for this remnant, though, does
not preclude its detection at shorter or longer wavelengths than the
{\it Chandra} operational energy bands.  In particular, this research
suggests that imaging of the remnant at those wavelengths at which
emission from ejecta material is relatively low may reveal such an
object.  Recently, \cite{C09} presented the detection of a radio
pulsar in association with Galactic SNR G315.9-0.0.
This remnant displays a similar emission ``spoke,'' extending
perpendicularly outward from the remnant, with the radio pulsar
detected at the outermost tip of \spoke.  The remnant E0102 and
G315.9-0.0 should not be directly compared because of morphological
differences and different estimated ages, but the results of
\cite{C09} and this research suggests that a more extensive review at
radio wavelengths could provide insight into the nature of a compact
object for this remnant.  Furthermore, the recent result of 16
galactic pulsars discovered with the {\it Fermi Gamma Ray Telescope}
\citep{A09} suggests that imaging of the remnant at higher energies
than the {\it Chandra} bands may be fruitful. In addition, a high-
resolution timing study of the center of the {\it TSR} could detect a
pulsar.  At present, {\it Chandra} GO program 11500201 (PI
$\!\colon\!$ Petre) is approved for 20\,ks of imaging with HRC-I in
Spring 2010.  These data could be used in the assessment of temporal
variability of an, as of yet, undetected central pulsar.

\section{Conclusion}

We have combined 324\,ks of {\it Chandra} X-ray observations of the
SNR 1E0102.2---7219 in the SMC to produce the highest
signal-to-noise image of this young, oxygen-rich remnant.  In this
research, we attempt to identify a compact stellar object for this
remnant within a generous search area motivated by observations of
this remnant, Galactic SNRs, and compact stellar objects.  We cannot
confirm the presence of a compact object for this remnant, but are
able to provide the strongest upper limits on the flux, assuming that
the spectral behavior of a faint neutron star for this remnant is best
described by a blackbody or power-law spectrum energy distribution.

\section{Acknowledgments}

This research has made use of data obtained from the Chandra Source
Catalog, provided by the {\it Chandra} X-ray Center (CXC) as part of the
{\it Chandra} Data Archive.  This research was partly funded by the Arizona
State University---NASA Space Grant program.  M. Rutkowski recognizes the Harvard-SAO Summer Research Experience Program for
Undergraduates.  The authors thank an anonymous referee
for comments that significantly improved the manuscript.\newpage

%\begin{verbatim}
%\begin{thebibliography}

\newpage
\begin{deluxetable}{ccr}
\tabletypesize{\footnotesize}
\tablecaption{E0102 Calibration Observations with S3 Chip Events Providing at Least 1\arcsec Resolution}
 \tablewidth{0pt} 
\tablehead{\colhead{Year}&
\colhead{Exposure Time (ks)}&
\colhead{ObsIDs}} 
\startdata 
2000 & 35.00 & 1308, 1423, 1803 \\ 
2001 & 32.48 & 141, 1311, 2843, 2844 \\ 
2002 & 15.33 & 2850, 2851\\
2003 & 51.70 & 3519, 3544, 5123, 5251, 5252 \\ 
2004 & 55.12 & 5130, 5131, 6074, 6075\\
2005 & 26.75 & 6042, 6043 \\
2006 & 53.30 & 6758, 6759, 6765, 6766 \\ 
2007 & 20.98 & 8365 \\ 
2008 & 19.20 & 9694 \\
2009 & 14.57 & 10655, 10656  
\enddata
\label{s3tables.tab}
\end{deluxetable}

\begin{deluxetable}{ccccccccc}
\centering
\tabletypesize{\footnotesize}
\tablewidth{0pc} 
\tablecaption{Upper Limits to the Flux of a Cas--A--like Neutron Star in Remnant 1E0102.2-7219} 
\tablehead{ 
\colhead{Region\tablenotemark{1}} & 
\colhead{Count Rate\tablenotemark{2}} &
\colhead{Flux \tablenotemark{3}} & 
\colhead{Luminosity \tablenotemark{4}} & 
\colhead{$\frac{L_{x}}{L_{C}}$\tablenotemark{5}} & 
\colhead{} & 
\colhead{Flux} & 
\colhead{Luminosity} & 
\colhead{$\frac{L_{x}}{L_{C}}$}}
\startdata 
\colhead{} &
\colhead{} &
\multicolumn{3}{c}{Assume $\Gamma=$2.35} & 
\colhead{} & 
\multicolumn{3}{c}{Assume $\Gamma=$3.13} \\
\cline{3-5} \cline{7-9} \\
1 & 3.61 (--4) & 1.12 (--14) & 4.65 (33) & 0.62 & &  3.73 (--14) & 1.55 (34) & 0.35 \\
2 & 1.61 (--3) & 5.01 (--14) & 2.09 (34) & 2.81 & &  1.66 (--13) & 6.89 (34) & 1.60 \\ 
3 & 1.41 (--3) & 4.39 (--14) & 1.82 (34) & 2.46 & &  1.46 (--13) & 6.06 (34) & 1.41 \\
4 & 3.57 (--3) & 1.11 (--13) & 4.61 (34) & 6.22 & &  3.69 (--13) & 1.53 (35) & 3.56 \\
\colhead{} &
\colhead{} & 
\multicolumn{3}{c}{Assume kT=0.53} & 
\colhead{} & 
\multicolumn{3}{c}{Assume kT=0.49} \\
\cline{3-5} \cline{7-9} \
1 & 3.61 (--4) &  3.49 (--15) & 1.45 (32) & 0.85 & & 3.46 (--15) & 1.44 (33) & 0.72 \\ 
2 & 1.61 (--3) &  1.56 (--14) & 6.47 (33) & 3.81 & & 1.55 (--14) & 6.43 (33) & 3.21 \\ 
3 & 1.41 (--3) &  1.37 (--14) & 5.69 (33) & 3.34 & & 1.35 (--14) & 5.60 (33) & 2.80 \\ 
4 & 3.57 (--3) &  3.46 (--14) & 1.44 (34) & 8.45 & & 3.43 (--14) & 1.42 (34) & 7.12 \\ 
\enddata 
\tablecomments{(--k) is equivalent to $\times 10^{-k}$}
\tablenotetext{1}{For the location of each region, see Figure \ref{fig:E0102zoom}; $^2$ in counts s$^{-1}$; $^3$ Unabsorbed (0.1-10 keV) Flux [erg~s$^{-1}$~cm$^{-2}$] derived from Observed count rate, and calculated by PIMMS assuming Cycle 10 systematics for ACIS-S with no gratings; $^4$ Unabsorbed Luminosity  [erg s$^{-1}$]; $^5$ $L_{x}$ $\!\colon\!$ Estimate of the Unabsorbed Luminosity from previous column, $L_{C}$ $\!\colon\!$ Unabsorbed Luminosity determined by \cite{C01}.}
\label{tab:powerlawest}
\end{deluxetable} 

\newpage

\newpage
\begin{landscape}
\begin{deluxetable}{cccccccccc}
\tabletypesize{\tiny}
\tablecolumns{10} 
\tablewidth{0pc} 
\tablecaption{Source Detections with {\sc wavdetect} in E0102} 
\tablehead{\colhead{Band\tablenotemark{1}} & \colhead{Search Scale\tablenotemark{2}} & \colhead{PSF Ratio\tablenotemark{3}} & \colhead{Significance\tablenotemark{4}} & \colhead{} & \colhead{$N_{rem}$\tablenotemark{5}} &  \colhead{PSF Ratio\tablenotemark{6}} & \colhead{Min. PSF Ratio\tablenotemark{7}} & \colhead{Significance\tablenotemark{6}} & \colhead{Max. Significance\tablenotemark{8}} }
\startdata 
\colhead{} & \colhead{} & \multicolumn{2}{c}{Nearest Detection\tablenotemark{9}} & \colhead{} & \multicolumn{5}{c}{Remnant Detections} \\
\cline{3-4} \cline{6-10}
\multirow{2}{*}{Broad} & Coarse & 5.72 & 176.4 & & 11 & 6.9 $\pm$ 2.1 & 3.41 & 544 $\pm$ 259 & 922 \\ 
& Fine & 1.98 & 98.5 & & 93 & 2.2 $\pm$ 0.8 & 1.6 & 31 $\pm$ 32 & 132 \\
\cline{1-10} 
\multirow{2}{*}{Soft} & Coarse & 5.73 & 159.9 & & 10 & 7.1 $\pm$ 1.8 & 3.46 & 556  $\pm$ 212 & 887 \\
& Fine & 1.95 & 48.2 & & 91 & 2.1 $\pm$ 0.8 & 1.4 & 31 $\pm$ 32 & 139 \\
 \cline{1-10} 
\multirow{2}{*}{Hard} & Coarse & 5.56 & 16.4 & & 15 & 6.0 $\pm$ 2.3 & 5.10 &  38 $\pm$ 19 & 76 \\
& Fine & 1.95 & 8.1 & & 31 & 2.2 $\pm$ 0.7 & 1.75 & 6 $\pm$ 2 & 10 \\
\cline{1-10} 
\multirow{2}{*}{Medium} & Coarse & 6.85 & 130.4 & & 13 & 6.9 $\pm$ 2.0 & 3.42 & 172 $\pm$ 74 & 302 \\
& Fine & 3.21 & 36.7 & & 76 & 2.4 $\pm$ 0.8 & 1.8 & 15 $\pm$ 10 & 45 \\
 \cline{1-10} 
\multirow{2}{*}{``Bin 1''} & Coarse & 7.17 & 11.5 & & 11 & 7.1  $\pm$ 2.01 & 3.09  & 51 $\pm$ 17 & 79 \\
& Fine & 1.96 & 6.4 & & 26 & 1.8 $\pm$ 0.9 & 1.8 &  6 $\pm$ 3 & 14 \\
\cline{1-10} 
\multirow{2}{*}{``Bin 2''} & Coarse & 3.11 & 47.7 & & 10 & 7.2  $\pm$ 2.3 & {\it 3.11} & 187 $\pm$ 71 & 308 \\
& Fine & 3.13 & 29.6 & & 67 & 2.2 $\pm$ 0.8 & 1.8 & 15 $\pm$ 11 & 46\\
 \cline{1-10} 
\multirow{2}{*}{``Bin 3''} & Coarse & 6.83 & 44.4 & & 11 & 6.9  $\pm$ 1.9 & 4.74  & 107 $\pm$ 47 & 172 \\
& Fine & 3.20 & 23.3 & & 63 & 2.1 $\pm$ 0.9 & 1.4 & 10 $\pm$ 6 & 30
\enddata 
\tablenotetext{1}{See Sections \ref{sec:twosig} and \ref{sec:pointsources} for energy band definitions.}
\tablenotetext{2}{``Fine'' $\equiv$ {\sc wavdetect} ``scale'' parameter $=$ ``1 2''; ``Coarse'' $\equiv$ {\sc wavdetect} parameter ``scale'' $=$ ``4 8 12''.}
\tablenotetext{3}{{\sc wavdetect} parameter PSFRATIO which can be used identifying extended sources; see Section \ref{sec:pointsources}.}
\tablenotetext{4}{{\sc wavdetect} parameter SRC\_SIGNIFICANCE; see Section \ref{sec:pointsources}.}
\tablenotetext{5}{Number of detections within $\sim$ 25$\farcs$ of the E0102 expansion center.}
\tablenotetext{6}{Mean, and 1$\sigma$ variance from mean, of {\sc wavdetect} parameter PSFRATIO for all detections ($N=$$N_{rem}$) in energy band.}
\tablenotetext{7}{Minimum PSFRATIO value for source detection interior to E0102.}
\tablenotetext{8}{Maximum SRC\_SIGNIFICANCE value for source detection interior to E0102.}
\tablenotetext{9}{Detection nearest, in proximity, to Region 4.}
\label{tab:wavdetect}
\end{deluxetable}
\end{landscape}

\newpage
\begin{landscape}
\begin{deluxetable}{cccccccccccccc}
\tabletypesize{\tiny}
%\rotate
\tablecolumns{14} 
\tablewidth{0pc} 
\tablecaption{Upper Flux Limits from Spectral Analysis to an E0102 point source$\colon$~Power Law Model \label{tab:fluxtab}}
\tablehead{ 
\colhead{Region} & 
\colhead{$F_{s}$\tablenotemark{\dag}} &
\colhead{$L_{s}$\tablenotemark{\dag\dag}} &
\colhead{$F_{h}$} &
\colhead{$L_{h}$} & 
\colhead{Min. $\chi^2_{\nu}$\,\,\tablenotemark{\ddag}} &
\colhead{} &
\colhead{$\Gamma$, Photon Index\tablenotemark{a}} &
%\colhead{Norm \tablenotemark{e}} & 
%\colhead{F$_{s}$} & 
%\colhead{F$_{h}$} & 
\colhead{$\Delta F_s$ \tablenotemark{b}} & 
\colhead{$F_s$ \tablenotemark{c}} & 
\colhead{$L_s$ \tablenotemark{d}} & 
\colhead{$\Delta F_h$} & 
\colhead{$F_s$} & 
\colhead{$L_h$}}
\startdata 
%\multicolumn{6}{c}{F$_{bkg,s} =$ 2.167 (-14), F$_{bkg,h}  =$ 2.433 (-13) ||| min $\chi^2  =$ 616.22 } \\
\colhead{} & \multicolumn{5}{c}{Background Fit} & \colhead{} & \multicolumn{7}{c}{Point Source Fit} \\
\cline{2-6} \cline{8-14}
\multirow{3}{*}{{\it p1}} & \multirow{3}{*}{2.167 (--14)} & \multirow{3}{*}{8.992 (33)} &  \multirow{3}{*}{2.433 (--13)} & \multirow{3}{*}{1.010 (35)} & \multirow{3}{*}{1.07} & & 1 & 9.600 (--16) &  2.263 (--14) & 9.391 (33)  & 3.800 (--15) & 2.471 (--13) & 1.025 (35) \\ 
& & & & & & & 3 & 9.000 (--16) & 2.257 (--14) & 9.366 (33)  & 2.000 (--16) &  2.435 (--13) & 1.010 (35) \\ 
& & & & & & & 5 & 3.600 (--16) & 2.203 (--14) & 9.142 (33) & 0.000 & 2.433 (--13) & 1.010 (35) \\
\cline{1-14}
\multirow{3}{*}{{\it  p2}} & \multirow{3}{*}{2.202 (--14)} & \multirow{3}{*}{9.138 (33)}  & \multirow{3}{*}{2.407 (--13)} & \multirow{3}{*}{9.988 (34)} & \multirow{3}{*}{1.05} & & 1 & 7.900 (--16) & 2.281 (--14) & 9.465 (33) & 3.100 (--15) & 2.438 (--13) & 1.011 (35) \\ 
& & & & & & & 3 & 7.300 (--16) & 2.275 (--14) & 9.441 (33) & 1.000 (--16) &  2.408 (--13) & 9.992 (34) \\ 
& & & & & & & 5 & -1.500 (--16) & 2.187 (--14) &  9.075 (33) & 8.900 (--15) &  2.496 (--13) & 1.036 (35) \\ 
% \cline{1-14}
% \\ %
\multirow{3}{*}{{\it  r1}} & \multirow{3}{*}{2.185 (--14)} & \multirow{3}{*}{9.067 (33)} & \multirow{3}{*}{2.443 (--13)} & \multirow{3}{*}{1.014 (35)} & \multirow{3}{*}{1.08} & & 1 & 8.200 (--16) & 2.267 (--14) & 9.407 (33) & 3.300 (--15) & 2.476 (--13) &  1.027 (35) \\ 
& & & & & & & 3 & 9.000 (--16) &  2.275 (--14) & 9.441 (33) & 2.000 (--16) &  2.445 (--13) & 1.015 (35)  \\ 
& & & & & & & 5 & 3.600 (--16) &  2.221 (--14) & 9.216 (33) & 0.000 &  2.443 (--13) & 1.014 (35)\\ 
\cline{1-14}
%\\ 
\multirow{3}{*}{{\it  r2}} & \multirow{3}{*}{2.222  (--14)} & \multirow{3}{*}{9.221 (33)} & \multirow{3}{*}{2.457 (--13)} & \multirow{3}{*}{1.020 (35)} & \multirow{3}{*}{1.05} & &  1 & 8.000 (--16) & 2.302 (--14) & 9.552 (33) & 3.100 (--15) &  2.488 (--13) & 1.032 (35) \\ 
& & & & & & & 3 & 7.900 (--16) &  2.301 (--14) & 9.548 (33) & 1.000 (--16) & 2.458 (--13) & 1.019 (35) \\ 
& & & & & & & 5 & 3.600 (--16) & 2.258 (--14) & 9.369 (33) & --1.000 (--16) & 2.456 (--13) & 1.019 (35) \\ 
\cline{1-14}
% \\ 
\multirow{3}{*}{{\it  r3}} & \multirow{3}{*}{2.222 (--14)} & \multirow{3}{*}{9.221 (33)} & \multirow{3}{*}{2.357 (--13)} & \multirow{3}{*}{9.781 (34)} & \multirow{3}{*}{1.06} & & 1 & 8.000 (--16) &  2.302 (--14) & 9.553 (33) & 3.200 (--15) &  2.389 (--13) & 9.913 (34) \\ 
& & & & & & & 3 & 9.000 (--16) & 2.312 (--14) & 9.594 (33) & 2.000 (--16) & 2.359 (--13) & 9.789 (34) \\ 
& & & & & & & 5 & 4.200 (--16) & 2.264 (--14) & 9.395 (33) & 0.00 &  2.357 (--13) & 9.780 (34) \\ 
 \cline{1-14}
\multirow{3}{*}{{\it  r5}} & \multirow{3}{*}{2.220  (--14)} & \multirow{3}{*}{9.212 (33)} & \multirow{3}{*}{2.435 (--13)} & \multirow{3}{*}{1.010 (35)} & \multirow{3}{*}{1.05} & & 1 &  8.000 (--16)  &  2.300 (--14) & 9.544 (33) & 3.200 (--15) & 2.467 (--13) & 1.023 (35) \\ 
& & & & & & & 3  & 8.900 (--16) & 2.309 (--14) & 9.581 (33)  & 2.000 (--16) &  2.437 (--13) & 1.011 (35)  \\  
& & & & & & & 5  & 3.000 (--16) &  2.250 (--14) & 9.336 (33) & 0.000  & 2.435 (--13) & 1.010 (35) \\ 
 \cline{1-14}
% \\ 
\multirow{3}{*}{{\it  r6}} & \multirow{3}{*}{2.213  (--14)} & \multirow{3}{*}{9.183 (33)} & \multirow{3}{*}{2.444 (--13)} & \multirow{3}{*}{1.014 (35)} & \multirow{3}{*}{1.05} & & 1 & 8.100 (--16) &  2.294 (--14) & 9.519 (33) & 3.200 (--15) & 2.476 (--13) & 1.027 (35) \\ 
& & & & & & & 3 & 8.300 (--16) &  2.296 (--14) & 9.527 (33) & 2.000 (--16) & 2.446 (--13)& 1.015 (35) \\ 
& & & & & & & 5 & 3.600 (--16) & 2.249 (--14) & 9.332 (33) & 0.000 &  2.444 (--13) & 1.014 (35) \\ 
 \cline{1-14}
% \\ 
\multirow{3}{*}{{\it  r7}} & \multirow{3}{*}{2.200  (--14)} & \multirow{3}{*}{9.129 (33)} & \multirow{3}{*}{2.339 (--13)} & \multirow{3}{*}{9.706 (34)} & \multirow{3}{*}{1.05} & & 1 &  8.100 (--16) & 2.281 (--14) & 9.465 (33)  & --7.000 (--15) &  2.269 (--13) & 9.416 (34) \\ 
& & & & & & & 3 &  8.400 (--16) &  2.284 (--14) & 9.477 (33) & --1.000  (--14) &  2.239 (--13) & 9.291 (34) \\ 
& & & & & & & 5 &  3.900 (--16) &  2.239 (--14) & 9.291 (33) & --1.030  (--14) &  2.236 (--13) & 9.278 (34) \\ 
\enddata 
\tablecomments{( -- k) is equivalent to $\times 10^{-k}$}
\tablenotetext{\dag}{$\colon$Modeled background flux ``s'' designates the soft X-ray band ( 0.5 -- 2.0\,keV ) and ``h'' designates the hard X-ray band ( 2.0 -- 8.0\,keV ) modeled flux (erg~s$^{-1}$~cm$^{-2}$)~---~$^{\dag\dag}$$\colon$Equivalent background luminosity, assuming distance to E0102 equal to 59\,kpc; ``s'' designates the soft X-ray band ( 0.5 -- 2.0\,keV ) and ``h'' designates the hard X-ray band ( 2.0 -- 8.0\,keV ) modeled flux (erg~s$^{-1}$~cm$^{-2}$)~---~$^{\ddag}$$\colon$Minimum $\chi^2_{\nu}$, $\nu$=577;  please see Section \ref{sec:specanalysis}~---~$^{a}$$\colon$ Assumed Power Law Index, $\Gamma$~---~$^{b}$$\colon$ the difference between the modeled unified background flux and the flux of the model point source, in energy band {\it x}~---~$^{c}$ the point source upper flux limit, in energy band {\it x}~---~$^{d}$$\colon$ the equivalent point source upper luminosity, in energy band {\it x}} 
%\label{tab:fluxtab}
\end{deluxetable}
\end{landscape}

\newpage
\begin{landscape}
\begin{deluxetable}{cccccccccccccc}
%\rotate
\tabletypesize{\tiny}
\tablecolumns{14} 
\tablewidth{0pc} 
\tablecaption{Upper Limit Flux for the Brightest Pixel Region\tablenotemark{\dag} in E0102} 
\tablehead{ \colhead{}
 & \colhead{$F_{s}$\tablenotemark{\dag\dag}} & \colhead{$L_s$\tablenotemark{\ddag}} & \colhead{$F_{h}$} & \colhead{$L_{h}$} & \colhead{Min. $\chi^2$\tablenotemark{\ddag\ddag}} & \colhead{} & \colhead{Fixed Parameter\tablenotemark{a}} & \colhead{$\Delta F_s$\tablenotemark{b}} & \colhead{$F_{s}$\tablenotemark{c}} & \colhead{$L_{s}$\tablenotemark{d}} & \colhead{$\Delta F_{h}$} & \colhead{$F_{h}$} & \colhead{$L_{h}$} }
\startdata 
\colhead{} & \multicolumn{5}{c}{Background Fit} & \colhead{} & \multicolumn{5}{c}{Point Source Fit} \\
\cline{2-6} \cline{8-14} \\
\multirow{3}{*}{Blackbody Fit} & \multirow{3}{*}{2.171 (--14)} & \multirow{3}{*}{9.009 (33)} & \multirow{3}{*}{2.309 (--13)} & \multirow{3}{*}{9.581 (34)} & \multirow{3}{*}{0.78} & & 0.25 & 0.163 (--14) & 2.334 (--14) & 9.685 (33) & 0.001 (--13) & 2.310 (--13) & 9.585 (34)\\
& & & & & & & 0.5 & 0.113 (--14) & 2.284 (--14) & 9.478 (33) & 0.008 (--13) & 2.317 (--13) & 9.615 (34)\\
& & & & & & & 1.0 & 0.061 (--14) & 2.232 (--14) & 9.262 (33) & 0.025 (--13) & 2.334 (--13) & 9.685 (34) \\
\cline{1-14} \\
\multirow{3}{*}{Power--Law Fit} & \multirow{3}{*}{2.199 (--14)} & \multirow{3}{*}{9.125 (33)}  & \multirow{3}{*}{2.442 (--13)} & \multirow{3}{*}{1.010 (35)}& \multirow{3}{*}{1.07} & & 1 & 7.900 (--16) & 2.278 (--14) & 9.453 (33)  & 3.100 (--15) & 2.473 (--13) & 1.026 (35) \\
& & & & & & & 3 & 7.600 (--15) & 2.959 (--14) & 1.228 (34) & 1.00 (--16) & 2.443 (--13) & 1.014 (35) \\
& & & & & & & 5 & 3.300 (--16) & 2.232 (--14) & 9.262 (33) & 0.00 & 2.442 (--13) & 1.013 (35) \\
\enddata 
\tablecomments{ ( --k ) is equivalent to $\times 10^{-k}$.}
\tablenotetext{\dag}{$\colon$Region 4, see Figure \ref{fig:E0102zoom}\@~---~$^{\dag\dag}$$\colon$Modeled Background Flux ``s'' designates the soft X-ray band ( 0.5 -- 2.0\,keV ) and ``h'' designates the hard X-ray band ( 2.0 -- 8.0\,keV ) modeled flux (erg~s$^{-1}$~cm$^{-2}$)~---~$^{\ddag}$$\colon$Equivalent Background Luminosity, assuming distance to E0102 equal to 59\,kpc; ``s'' designates the soft X-ray band ( 0.5 -- 2.0\,keV ) and ``h'' designates the hard X-ray band ( 2.0 -- 8.0\,keV ) modeled flux (erg~s$^{-1}$~cm$^{-2}$)~---~$^{\ddag\ddag}$$\colon$Minimum $\chi^2_{\nu}$, $\nu$=577; please see Section \ref{sec:specanalysis}~---~$^{a}$$\colon$Assumed model parameter, $\Gamma$ or kT$_{\mbox{eff}}$ (keV)~---~$^{b}$$\colon$$\Delta F_{x}$ is equal to difference between the modeled unified background flux and the flux of the model point source in energy band {\it x}~---~$^{c}$$\colon$ the point source upper flux limit, in energy band {\it x}~---~$^{d}$$\colon$ the point source upper luminosity limit, in energy band {\it x}} 
\label{tab:region4flux}
\end{deluxetable}
\end{landscape}

\newpage

\newpage

\end{document}